\documentclass[twoside]{LCWS13}
\usepackage[latin1]{inputenc}
\usepackage[dvips]{graphicx,epsfig,color}
\usepackage{wrapfig,rotating}
\usepackage{amssymb,amsmath,array}
\usepackage{cite}
\newcommand{\kakeru}{$\times$}

\begin{document}
\title{
%%%%   Paper title goes here  %%%%%%%%%%%%%%
Scintillator Strip ECAL Optimization} %% 
%***********************************************************************
% AUTHORS INFORMATION AREA
%***********************************************************************
%\author{Katsushige Kotera$^1$ and Adil Khan$^2$ on behalf of the CALICE collaboration
\author{Katsushige Kotera on behalf of the CALICE collaboration
% Optional short acknowledgment: remove next line if non-needed
%\thanks{This is an optional funding source acknowledgment.}
% DO NOT MODIFY THE FOLLOWING '\vspace' ARGUMENT
\vspace{.3cm}\\
% Addresses and institutions (remove "1- " in case of a single institution)
Shinshu University, Faculty of Science \\
Asahi-3-1-1, Matsumoto, Nagano, Japan
%% Remove the next three lines in case of a single institution
\vspace{.1cm}\\
%2- Kyungpook National University - Center For High Energy Physics Department \\
%Daegu - S - Korea\\
}
%%***********************************************************************
% END OF AUTHORS INFORMATION AREA
%***********************************************************************

\maketitle

\vspace{-7mm}
{\it Talk presented at the International Workshop on Future Linear Colliders (LCWS13), Tokyo, Japan, 11-15 November 2013.}
\vspace{8mm}
\begin{abstract}
The CALICE collaboration is developing a granular electromagnetic calorimeter using small scintillator strips for a future  linear collider experiment.
%Each scintillator strip is read out by using a Pixelated Photon Detector (PPD).
On developing of $\sim 10^7$ channel-ECAL in particle flow approach, CALICE is developing a technological prototype with 144 of 5\,$\times$\,45\,$\times$\,(1 - 2)\,mm$^3$ strips on each 180\,$\times$\,180\,mm$^2$ base board unit
in tandem with developing the design of scintillator strip and pixelated photon detector and their coupling after established the physics prototype which has required performance.
A method of event reconstruction in such ECAL is also developed. % including the way which uses square tile scintillator layers interleaved into strip layers.
\end{abstract}

\section{Introduction}

%%%%%%%%%%%%%%%%%%%%%%%%%%%%%%%%%%
%%%%%%%%%%%%%%%%%%%%%%%%%%%%%%%%%%%%%%
In the future linear collider experiments the particle flow approach (PFA) is one of the most potential method to have high performance jet physics.
For the PFA, a granular sampling electromagnetic calorimeter (ECAL) is mandatory \cite{MarkT}. 
%%%%%%%%%%%%%%%%%%%%%%%%%%%%%%%%%%%%%%%%
%%%%%%%%%%%%%%%%%%%%%%%%%%%%%%%%%%%%%%%%
With 5\,\kakeru\,5\,mm$^2$ lateral segmentation typically required to the ECAL in the International Large Detector (ILD) \cite{DBD} for the International Linear Collider (ILC), 
the number of channels becomes $10^8$, while it can be reduced to $10^7$ with scintillator strips of 45\,\kakeru\,5\,mm$^2$  lateral shape 
aligned orthogonally in successive layers.

\begin{figure}[!h]
\centerline{\includegraphics[width=0.75 \columnwidth]{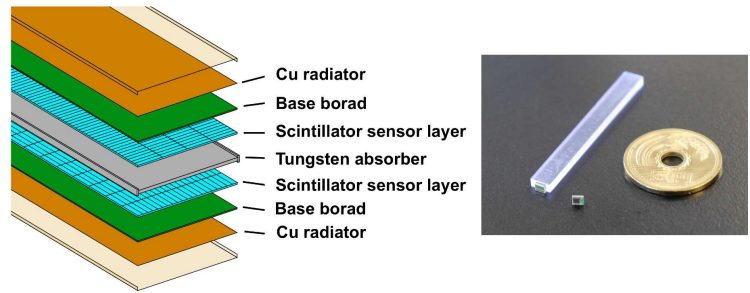}}
\caption{{\it Left}; A pair of scintillator layers designed for ILD ScECAL. Pairs of 10 - 15 are inserted into the shelf structured tungsten plates simultaneously work as the absorber plates. {\it Right}; a 45\,\kakeru\,5\,\kakeru\,2\,mm$^3$ scintillator strip and a PPD (1600-pixel MPPC produced by Hamamatsu Photonics KK \cite{HPK})}\label{fig:ScECALInILD}
%\end{wrapfigure}
\end{figure}

The CALICE collaboration is developing such an ECAL using scintillator strips read out with Pixelated Photon Detectors (PPDs) as the sensitive layers interleaved with tungsten plates as the absorber layers \cite{LOI}. 
Figure \ref{fig:ScECALInILD} shows the design of two sensitive layers of such a scintillator electromagnetic calorimeter (ScECAL) with an absorber layer.
This, a slab, is inserted into a shelf structure of $\sim$ 15 other tungsten plates constructing the barrel and end-caps of ILD-ECAL. Details are shown in some publications \cite{DBD, CALICEDOC}. 
A prototype module of ScECAL has shown a expected performance of the single particle energy resolution in the quadratic parametrization of stochastic and constant terms of  
12.9$\pm$0.1(stat.)$\pm$0.4(syst.)\% and 1.2$\pm$0.1(stat.)$^{+0.4}_{-1.2}$(syst.)\%, respectively to the electrons with energy range of 2 - 32 GeV \cite{CAN16b}.

As the next step of ECAL development, the readout system should be embedded in between ECAL layers to reduce the dead volume comes from the large amount of cables of $\sim 10^7 - 10^8$ channels.
This requirers the compactness of readout system increasing the density of electronics on the baseboard: 144 channels/180\,\kakeru\,180\,mm$^2$.
In this paper the technical prototype developed such thinner and dense readout board is introduced in the successive section.

At the time of development of physics prototype, it was required to have 10\,\kakeru\,10\,mm$^2$ segmentation.
Therefore, the strips were allowed to have 10\,mm width, and each scintillator strip then was available to be read out by using the wavelength shifting (WLS) fiber to collect the scintillation photons. % in the strip.
However, according to a later study \cite{LOI} the requirement was changed to 5\,\kakeru\,5\,mm$^2$ segmentation for the best PFA performance. 
Additional to this fact, thickness of scintillator strips are required to be thinner since we can reduce the cluster radius with thinner scintillator strips. 
Therefore, the CALICE ScECAL group made an effort to establish the scintillator-PPD unit which have enough photon yield and uniform response without WLS fiber. 
Section \ref{sec:scintillatorMPPC} shows the development of scintillator-PPD unit.
In the current technological prototype the number of pixels of PPD is 1,600. This is not enough when we consider the large energy clusters, since the PPD has the saturation phenomena.
Hamamatsu Photonics KK \cite{HPK} developed  a 10,000 pixel MPPC, a name of product, in 2013 and it will be discuss in section \ref{sec:MPPC}.
To extract the fine segmentation performance with ScECAL, a special event-reconstruction method is required and it has been developed.
In section \ref{sec:SSA} such reconstruction method, called the strip splitting algorithm (SSA) is shown and discussed,
including the way where we use tile layers between strip layers.
Finally we summize the optimization of ScECAL in section \ref{sec:summary}.
 
%%%%%%%%%%%%%%%%%%%%%%%%%%%%%%%%%%
%%%%%%%%%%%%%%%%%%%%%%%%%%%%%%%%%%%%%%
\section{Technological prototype}\label{sec:technologicalPrototype}
\begin{figure}[!h]
\centerline{\includegraphics[width=1.2 \columnwidth]{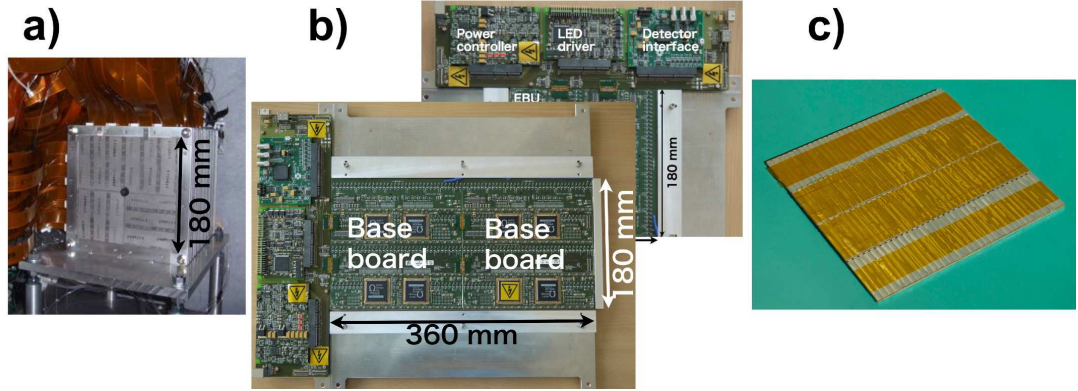}}
\caption{
a) ScECAL physics prototype tested at Fermilab 2008 and 2009. b) Two-layer ScECAL technological prototype tested at DESY 2013. First layer has two EBU (ECAL baseboard unit) and
the second layer behind the photo of the first layer has one EBU. c) The scintillator strips of 144 channels hermetically enveloped in reflector foil on the other side of EBU.
}\label{fig:EBU}
%\end{wrapfigure}
\end{figure}

%%%%%%%%%%%%%%%%%%%%%%%%%%%%%%%%%%
%%%%%%%%%%%%%%%%%%%%%%%%%%%%%%%%%%%%%%
Figure\,\ref{fig:EBU}a shows the physics prototype tested at Felmilab 2008 and 2009, and Fig.\,\ref{fig:EBU}b shows the two-layer technological prototype tested at DESY on July 2013.
The readout cable in Fig.\,\ref{fig:EBU}a and 
readout electronics out of the photo of the physics prototype were embedded on the ECAL base board units (EBU) of the technological prototype.
 Each EBU has 144 scintillator strips on the other side shown in Fig.\,\ref{fig:EBU}c.
 The size of scintillator strip is 43.4\,\kakeru\,4.9\,\kakeru\,2\,mm$^3$ and the size of PPD sensor area is 1.0\,\kakeru\,1.0\,mm$^2$ enveloped into 2.45\,\kakeru\,1.9\,\kakeru\,0.85\,mm$^3$ plastic package.
 The ladder of 36 strips individually-hermetically enveloped into reflector foil make a low of 180\,\kakeru\,43.4\,mm$^2$ scintillator plate, and four of such scintillator plates fill a surface of 180\,\kakeru\,180\,mm$^2$ EBU. 
 
 An EBU has four ASICs, SPIROC2b developed by OMEGA group\,\cite{SPIROC} mounted on the other side of scintillators.
 An ASIC reads out the analog signals from 36 PPDs and provides bias voltages of individual PPDs. 
 The all ASICs on a slab are driven by using a director interface board (DIF).
 This DIF has an FPGA controlling parameters of ASIC: bias voltages of individual channels, timing delay of data taking, threshold of auto-trigger function, amplitude of amplifiers, and so on.
 An EBU has LED on each channel for the gain monitoring. The power supply of the LED through the other board also controlled by DIF.
%%%%%%%%%%%%%%%%%%%%%%%%%% 

\begin{figure}[!h]
\centerline{\includegraphics[width=0.4 \columnwidth]{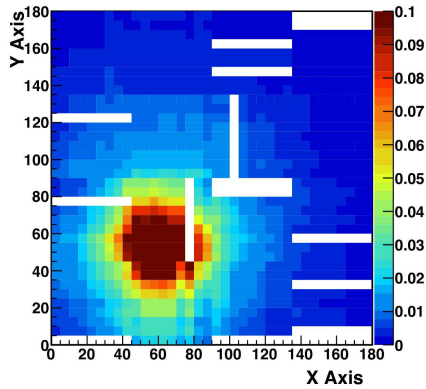}}
\caption{
Reconstructed energy deposited of 3 GeV electron beams on the two layer ScECAL technological prototype.
}\label{fig:protoSSA3GeV}
%\end{wrapfigure}
\end{figure}

 Each of two layers were aligned having strip direction orthogonally with respect to the other layer.
 Therefore, 5\,\kakeru\,5\,mm$^2$ segmentation reconstruction could be available to the electron shower events by using SSA\,\cite{SSA}.
 Figure\,\ref{fig:protoSSA3GeV} shows typical example of such reconstruction of events, where the intensity of averaged energy deposited in reconstructed 5\,\kakeru\,5\,mm$^2$ cells are indicated in colors.

 Details of performance of the technological prototype was discussed in the other talk of this LCSW by T.~Ogawa \cite{ogawa}.
 
 %%%%%%%%%%%%%%%%%%%%%%%%%%%%%%%%%%
%%%%%%%%%%%%%%%%%%%%%%%%%%%%%%%%%%%%%%
 \section{Scintillator-PPD unit} \label{sec:scintillatorMPPC}
 %%%%%%%%%%%%%%%%%%%%%%%%%%%%%%%%%%
%%%%%%%%%%%%%%%%%%%%%%%%%%%%%%%%%%%%%%
 Scintillator-PPD unit is the core part of ScECAL.
 In the technological prototype  
a 1,600 pixel MPPC is directly  coupled on one of the cut edges of each scintillator strip hermetically enveloped in a cut of reflector foil.
 % 43.4\,\kakeru\,4.9\,\kakeru\,2\,mm$^3$ is used with 1,600 pixel MPPC which is a product of PPD provided by Hamamatsu.
 Although this unit has mostly satisfied the requirements as the ILD-ECAL,
 some issues to be improved exist.
 In this section  we discuss status of development on those issues; thickness of scintillator strip, shape of PPD sensitive area, design of combination of a scintillator and a PPD, and the reflector on a scintillator surface.
 
 %%%%%%%%%%%%%%%%%%%%%%%%%%%%%%%%%%
%%%%%%%%%%%%%%%%%%%%%%%%%%%%%%%%%%%%%%
 \subsection{Thickness of scintillator strip}
\begin{figure}[!h]
\centerline{\includegraphics[width=0.4 \columnwidth]{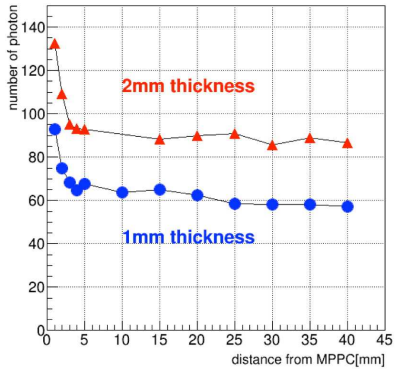}}
\caption{
Simulated photon yields as functions of distance between position where MIP like particles injected in and PPD, assuming that
the photon detection efficiency is 100\%. 
}\label{fig:onemmScinti}
%\end{wrapfigure}
\end{figure}

 Although the technological prototype was constructed with 2 mm thick scintillator strips, thinner scintillator allows us to reduce the cluster radius,
 although the reduced sampling ratio can be cause of the degradation of the intrinsic energy resolution for the single particles\footnote{
 PFA requires particle separations rather than the intrinsic energy resolution of the single particles. 
 }. 
 Additionally, thinner ECAL with thinner scintillator strips allows us to reduce the cost of ILD.
 According to the simulation result shown in Fig.\,\ref{fig:onemmScinti}, 
 the photon yield with 1 mm thick scintillator decreases by 2/3 of the case with 2 mm thick scintillator which has 6 - 7 p.e./MIP from the result of test beam at DESY \cite{ogawa} and 
 7 p.e./MIP is optimized number for the ScECAL acceding to our study.
 Optimization of design of scintillator strip will  be taken considering this fact.
 
\subsection{Shape of PPD}
\begin{figure}[!h]
\centerline{\includegraphics[width=0.7 \columnwidth]{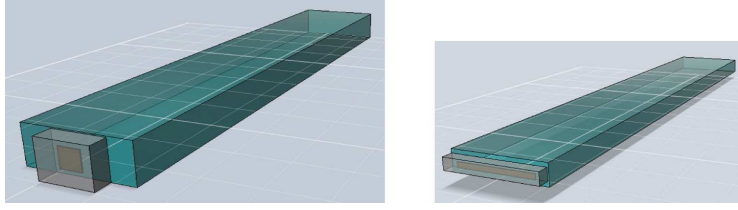}}
\caption{{\it Left}; A 
schematic of 1\,$\times$\,1\,mm$^2$ sensitive area PPD coupled on 2\,mm thick $\times$\,5\,mm width scintillator, and 
 {\it Right}; 
  0.25\,$\times$\,4\,mm$^2$ sensitive area PPD coupled on 1\,mm thick $\times$\,5\,mm width scintillator
 )}\label{fig:shapeMPPC}
%\end{wrapfigure}
\end{figure}

Figure\,\ref{fig:shapeMPPC}, {\it left} shows the current design of scintillator strip and PPD in the technological prototype which discussed in section \ref{sec:technologicalPrototype}.
With this combination, the response is totally flat but a steep peak exists right in front of PPD.
With the PPD, size of 0.25\,\kakeru\,4\,mm$^2$ shown in Fig.\,\ref{fig:shapeMPPC}, {\it right}, we can expect that the steep peak is moderated .
Therefore, we discussed with Hamamatsu Photonics KK to develop such MPPCs and confirmed that there was no technological difficulty to produce such MPPC.

\subsection{Combination of scintillator strip and PPD}
\begin{figure}[!h]
\centerline{\includegraphics[width=0.8 \columnwidth]{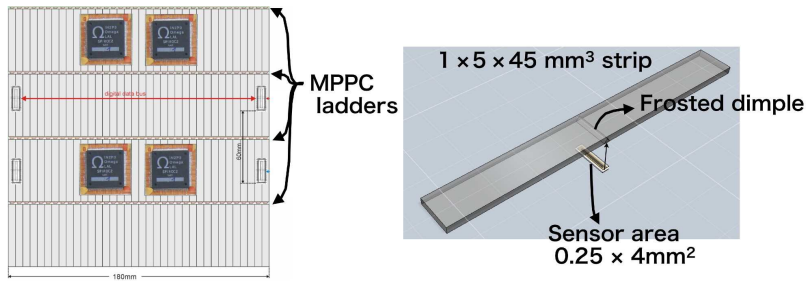}}
\caption{{\it Left}; An alignment of 144 scintillator strip on an EBU showing four dead volume line due to the PPDs. 
{\it Right}; a schematic of 0.25\,$\times$\,4\,mm$^2$ sensitive area PPD under a scintillator strip. )}\label{fig:centerReadOut}
%\end{wrapfigure}
\end{figure}

With the combination of scintillator-PPD shown in Fig.\,\ref{fig:shapeMPPC}, dead volume due to the PPD thickness occupies 2\% of lateral area of ECAL as shown in Fig.\,\ref{fig:centerReadOut} {\it left}.
Therefore, we try to read out the photons in the strips from the bottom of each strip as an example in Fig.\,\ref{fig:centerReadOut}, {\it right}. 
Further study of the light collection method with this connection is ongoing. 

\begin{figure}[!h]
\centerline{\includegraphics[width=0.9 \columnwidth]{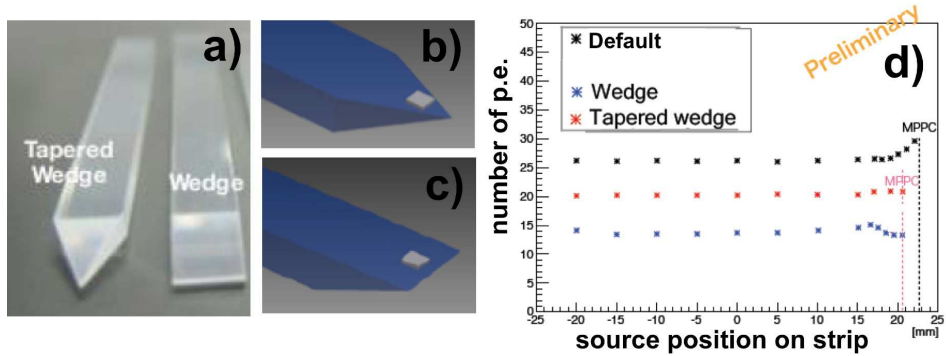}}
\caption{a); 
Two designs of scintillator strip with tapered wedge and not tapered wedge strips.
b) and c); PPD position on the bottom of wedge.
d); 
 The number of photo-electrons detected with two designs of scintillator strip and default design of strip as function of the position of the source of electrons ($^{90}$Sr) along the center line of 
  longitudinal direction of the scintillator strip, measured from an origin at the center of strip.}\label{fig:wedge_tapered}
%\end{wrapfigure}
\end{figure}

One of the potential ways to readout efficiently from the bottom is to make a wedge shape on one of the cut edges of strip as shown in Fig.\,\ref{fig:wedge_tapered}a.
Figure\,\ref{fig:wedge_tapered}b and \ref{fig:wedge_tapered}c show the position of 1\,\kakeru\,1\,mm$^2$ sensitive area of PPD on the bottom of wedge. 
A set of experimental measurements of such designs including the strip which has tapered sides of wedge has been done and the results are shown in \ref{fig:wedge_tapered}, d).
With the tapered wedge strip, steep peak of response is removed and photon yield is higher than the case without tapered sides\footnote{
The number of p.e. in Fig.\,\ref{fig:wedge_tapered} {\it right} is three themes larger than the number measured with the technological prototype measured with electron beams at DESY.
After this conference, we confirmed empirically that the differences of reflector foil, scintillator material, source of incidental events; $^{90}$Sr and mip like e$^+$,
 and observables; mean of distribution of p.e. and the most probable values, explain the main parts of differences. However those are not enough as the reasons and further investigation is ongoing.
}
 indicating high potential of this design in the ScECAL.

 More detail were discussed in  a talk of this conference by S.~Ieki \cite{ieki}.

\subsection{Reflection on the scintillator surface}
\begin{figure}[!h]
\centerline{\includegraphics[width=0.35 \columnwidth]{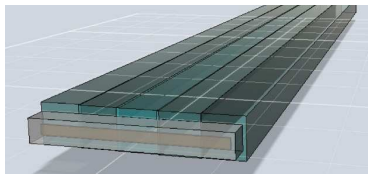}}
\caption{
A schematic of scintillator strip made of bundled five scintillation fibers which have clad surface to keep scintillation photons in the scintillator.
A  0.25\,\kakeru\,4\,mm$^2$ PPD is coupled.}\label{fig:fiveFibers}
%\end{wrapfigure}
\end{figure}

Current design of method to 
make complete reflection on the scintillator surface is to envelop the scintillator in a cut of reflector foil provided by KIMOTO Co., Ltd \cite{kimono}.
This foil not only works well to increase reflection but also works to prevent the optical cross-talk between scintillator strips.
However, the folding method and envelop procedure add a little complex to the construction of ECAL.
Therefore, some metal spattering methods are considered while we know the total reflection function is mandatory to have good reflection and response uniformity.
For example thick silver alloy, or plastic base layer outside of spattering.  
Additional design is that five of 1\,\kakeru\,1\,mm$^2$ cladding scintillator fibers are bundled as a 5\,\kakeru\,1\,mm$^2$ cross-section scintillator plates as shown in Fig.\,\ref{fig:fiveFibers}.
Studies are on going.

\subsection{PPD in ScECAL }\label{sec:MPPC}
\begin{figure}[!h]
\centerline{\includegraphics[width=0.4 \columnwidth]{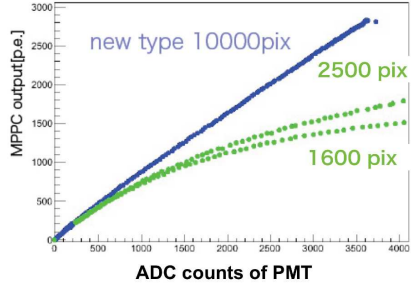}}
\caption{
The number of photo-electrons of several MPPCs.
The horizontal axis shows corresponding incident photons measured  by using a photo-multiplier. 
}\label{fig:mppcResponse}
%\end{wrapfigure}
\end{figure}
A PPD has a saturation phenomenon due to its finite number of pixels.
According to our estimation considering of $e^+e^-\to e^+e^-$ events at $\sqrt{s} = 500$\,GeV, we need 15,000 - 18,000 pixel PPD\footnote{
According to our Monte Carlo simulation study, the maximum energy deposited at a scintillator strip is 2,500 electrons, when we have 250 GeV electron.
Since 6 - 7  p.e./mip is the photon yield of the current design,
this corresponding to 15,000 - 18,000 pixels.}. 

Hamamatsu Photonics KK developed a new technology with metal register instead of polysilicon register in last year, and
this technology with other improvements leads them to succeed to produce 10,000 pixel PPD.
Therefore, the PPD development is close to the goal.
Figure\,\ref{fig:mppcResponse} shows the response curves of a several types of MPPC.
The saturation phenomenon of 10,000 pixel MPPC is very much moderated than others.
Study of saturation phenomenon with further amount of photons are ongoing.

\subsection{Summary of status of scintillator-PPD unit}
We introduced some designs and ideas of scintillator-PPD units.
After this conference, we are entering the stage converging those designs with some experiments in laboratories and some test beams.

\section{Reconstruction method}\label{sec:SSA}
 
 An algorithm to reconstruct strip ScECAL has been developed \cite{SSA}.
 In the algorithm the total energy deposited on a strip, $E_{\mathrm{strip}}$ 
 is split into virtual square cells where the strip is divided by its width along its length; a 45\,\kakeru\,5\,mm$^2$ strip is split into nine 5\,\kakeru\,5\,mm$^2$ cells. 
 %and the square is expanded a little when the length is not right multiple of width.
%As shown in Fig.\,\ref{fig:SSAexplanation} t
The energy is split conservatively according to the weights estimated by using the energy deposited on the strips in immediately neighboring layers, having intersection with the strip being considered, when seen from the interaction point of ILD.
The energy deposited on the virtual cell $k$ is estimated as the following:
\begin{equation}
	E_{\mathrm{virtual}}^k = E_{\mathrm{strip}}\cdot \Sigma_i E_{\mathrm{neighbor}}^{\{i,k\}}/ \Sigma_i E_{\mathrm{neighbor}}^i,
\end{equation}
where, $k$ is the index of the virtual cell within the strip and $i$ is the index of neighbouring intersecting strips, and $E_{\mathrm{neighbor}}^{\{i,k\}}$ is the energy deposited in the strip $i$ 
having intersection  in the range of virtual cell $k$.

\begin{figure}[!h]
\centerline{\includegraphics[width=0.7 \columnwidth]{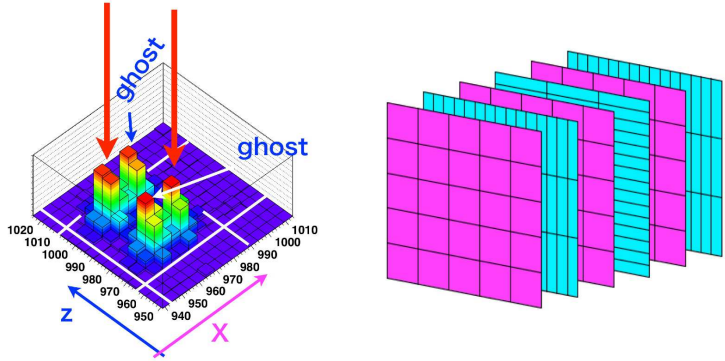}}
\caption{{\it Left}; Schematic of two particles simultaneously come into a square area enclosed with length of strips. 
Two excess clusters (ghosts) occur on the orthogonal position of true positions.
{\it Right}; Schematic of model of ScECAL layers alternately replaced strip layers with tile layers.}\label{fig:ghost}
%\end{wrapfigure}
\end{figure}

In an idea to prevent faked clusters due to the twofold ambiguity, when two particles simultaneously come into the square 
area enclosed in the strip length as shown in Fig.\,\ref{fig:ghost}, {\it left},
strip layers are alternately replaced with square tile layers as shown in Fig.\,\ref{fig:ghost}, {\it right}.
When the size of square tiles are larger than the virtual cells of strip cells, the square tiles are sprit in the virtual cells by using 
information of energy deposited of neighboring strip cells in the same way above, and the energies on strips then 
are split by using information of virtual cells originally comes from the tile cells.
The energy and position of all virtual cells are analyzed by a particle flow algorithm \cite{MarkT}.

\begin{figure}[!h]
\centerline{\includegraphics[width=0.5 \columnwidth]{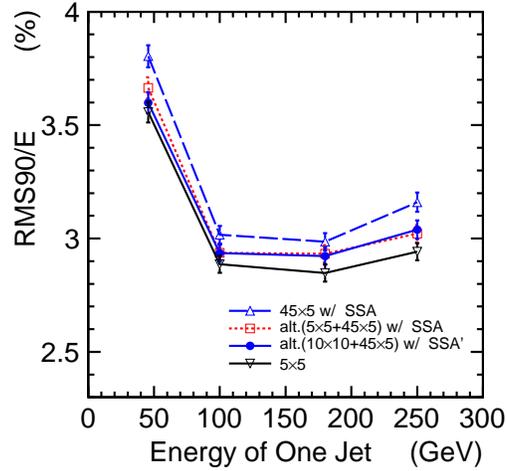}}
\caption{Jet energy resolutions as a function of energy of incident jet, $e^+ e^- \to u,d,s$ }\label{fig:JER}
%\end{wrapfigure}
\end{figure}

Figure\,\ref{fig:JER} shows simulation results of jet energy resolution as a function of the energy of jet incident into an ILD model.
The jet energy resolution with ScECAL of 45\,$\times$\,5\,mm$^2$ strips is close to the jet energy resolution with ScECAL of 5\,$\times$\,5\,mm$^2$ square tiles.
Difference between them is only $<$ 0.25\%.
Additionally, with ScECAL when we alternately replaced strip layers with 10\,$\times$\,10\,mm$^2$ square tile layers, ScECAL has the same jet energy resolution as the 5\,$\times$\,5\,mm$^2$ tile ScECAL has 
at the smaller jet energy below than 100 GeV, and only 0.1\% degrades at  energy above 100 GeV, indicating that the insertion of the tile layers are promising.

\section{Summary}\label{sec:summary}

The CALICE collaboration is developing a scintillator strip based ECAL for ILC detector.
Phase of research and developing was moved from the physics prototype to the technological prototype which are considering to implement into the real ILD detector:
the readout electronics is embedded between each sensitive layers outputing the digitized data of the energy deposited of each channel.
With two layer technological prototype, successfully reconstructed 5\,$\times$\,5\,mm$^2$ segmentation was demonstrated.
Although the current design has performance close to the goal, there exist some further potential to be optimized on the scintillator-PPD units:
thickness of scintillator, geometry of sensitive area of PPD, method to read out scintillation photon with PPD, easier method to increase reflection on the scintillator surface, and the 
PPD of the large number of pixels.
Additionally, we are planing to insert tile layers in between strip layers, and promising effect of such layers are confirmed. 
We will fix the design to some of the best design in near future.

\section{Acknowledgments}
The authors gratefully thank to the CALICE group for supporting development of ScECAL and for useful discussions, and thank to the ILD group for providing the software tools to study ScECAL in ILD.
This study is supported in part by the Grant-in-Aid for Specially Promoted Research No.~23000002 of the Japan Society for Promotion of Science (JSPS)
and 
by the Grant-in-Aid for Scientific Research on Innovative Areas No.~24104503 of the JSPS.

% ****************************************************************************
% BIBLIOGRAPHY AREA
% ****************************************************************************

\begin{footnotesize}
% IF YOU DO NOT USE BIBTEX, USE THE FOLLOWING SAMPLE SCHEME FOR THE REFERENCES
% ----------------------------------------------------------------------------

% ----------------------------------------------------------------------------

\end{footnotesize}

% ****************************************************************************
% END OF BIBLIOGRAPHY AREA
% ****************************************************************************


\begin{thebibliography}{99}
\bibitem{SSA} K.~Kotera {\it et al}, in~a~few~days~we~are~submit~a~paper~to~the~arxiv, and~replace~this~with~the~arxiv~number.
\end{thebibliography}
\end{document}